\documentclass[10pt,twocolumn,3p]{elsarticle}

\usepackage[utf8]{inputenc}
\usepackage[]{hyperref}
\usepackage[]{graphicx}

\usepackage{amsmath,amssymb}

\usepackage{miller}
\newcommand{\etal}{\textit{et al}.}
\newcommand{\ie}{\textit{i.e.}}
\newcommand{\insitu}{\textit{in situ}}
\newcommand{\Insitu}{\textit{In situ}}
\newcommand{\postmortem}{\textit{post mortem}}
\newcommand{\Postmortem}{\textit{Post mortem}}

\newcommand{\degree}{$^{\circ}$}

\journal{Acta Materialia}
\begin{document}

\begin{frontmatter}

\title{Mobility of $\langle c+a \rangle$ dislocations in zirconium}

\author[SRMP]{Thomas Soyez}
\author[CEMES]{Daniel Caillard}
\author[SRMA]{Fabien Onimus}
\author[SRMP]{Emmanuel Clouet\corref{CA}}
\cortext[CA]{Corresponding author}
\ead{emmanuel.clouet@cea.fr}

\address[SRMP]{Université Paris-Saclay, CEA, Service de Recherches de Métallurgie Physique, F-91191 Gif-sur-Yvette, France}
\address[CEMES]{CEMES-CNRS, 29 rue Jeanne Marvig, BP 94347, F-31055 Toulouse, France}
\address[SRMA]{Université Paris-Saclay, CEA, Service de Recherches Métallurgiques Appliquées, F-91191 Gif-sur-Yvette, France}

\begin{abstract}
	Plasticity in hexagonal close-packed zirconium is mainly controlled by the glide of dislocations 
	with $1/3\,\hkl<1-210>$ Burgers vectors.  As these dislocations cannot accommodate deformation
	in the $\hkl[0001]$ direction, twinning or glide of \hkl<c+a> dislocations, 
	\ie{} dislocations with $1/3\,\hkl<1-213>$ Burgers vector, have to be activated. 
	We have performed \insitu{} straining experiments in a transmission electron microscope 
	to study the glide of \hkl<c+a> dislocations in two different zirconium samples, 
	pure zirconium and Zircaloy-4, at room temperature. 
	These experiments show that \hkl<c+a> dislocations exclusively glide 
	in first-order pyramidal planes with cross-slip being activated.
	A much stronger lattice friction is opposing the glide of \hkl<c+a> dislocations 
	when their orientation corresponds to the \hkl<a> direction 
	defined by the intersection of their glide plane with the basal plane.  
	This results in long dislocations straightened along \hkl<a> 
	which glide either viscously or jerkily.
	This \hkl<a> direction governs the motion of segments with other orientations,
	whose shape is merely driven by the minimization of the line tension. 
	The friction due to solute atoms is also discussed.
\end{abstract}


\end{frontmatter}

\section{Introduction}

Zirconium alloys are used in the nuclear industry as fuel cladding tubes and structural components in light and heavy water reactors \cite{Motta2017}.
Like in other hexagonal close-packed (hcp) metals, 
plastic activity strongly varies between the different possible slip systems in zirconium.
The principal slip system is controlled by glide of \hkl<a> dislocations, \ie{} dislocations with $1/3\,\hkl<1-210>$ Burgers vectors, in the \hkl{10-10} prismatic planes
\cite{Caillard2003,Clouet2015}.
These \hkl<a> dislocations can also cross-slip above room temperatures to glide in the \hkl(0001) basal and \hkl{10-11} first-order pyramidal planes \cite{Caillard2018}. 
However all these deformation modes only accommodate deformation along \hkl<a> directions 
and other mechanisms, either twinning or glide of dislocations with a \hkl<c> component,
are needed to allow for a deformation along the $\hkl<c>=\hkl[0001]$ axis of the hcp crystal.

Both twinning and glide of \hkl<c+a> dislocations, \ie{} dislocations with $1/3\,\hkl<11-23>$ Burgers vectors, are active in zirconium
\cite{Tenckhoff1972,Jensen1972,Akhtar1973,Woo1979,Holt1987,Merle1987,Numakura1991,McCabe2006,Long2015a,Long2015b,Long2017},
with $\hkl<c+a>$ slip becoming more active when the temperature increases.
Depending of the hcp metals, these \hkl<c+a> dislocations can glide in the first-order \hkl{10-11} 
or second-order \hkl{-2112} pyramidal planes.
Almost all transmission electron microscopy (TEM) observations have shown, by slip traces analysis \cite{Akhtar1973} 
or stereography \cite{Woo1979,Merle1987,Numakura1991,Long2015a,Long2015b,Long2017,Long2018}, 
that \hkl<c+a> dislocations glide in first-order pyramidal planes in Zr alloys,
with possible cross-slip between two pyramidal planes sharing the same \hkl<c+a> direction \cite{Long2015b,Long2017,Long2018}.
Scanning electron microscopy of slip traces on bended micro-cantilever well oriented to activate only \hkl<c+a> slip
also concluded that \hkl<c+a> dislocation glide in first-order pyramidal planes, 
without any trace of slip in second-order pyramidal planes, 
both for the compressive and tensile parts of the cantilevers \cite{Gong2015}.
Only Long \etal{} \cite{Long2018} have recently evidenced that a minority of \hkl<c+a> can also be found in the second order pyramidal plane in a Zr-2.5Nb alloy. 
Besides, some of these TEM observations indicate that \hkl<c+a> dislocations have a tendency to straighten 
in the direction corresponding to the intersection of the pyramidal glide plane with the basal plane \cite{Woo1979,Numakura1991,Long2017,Long2018},
a feature which is not specific to zirconium and can be found in other hcp metals, like Mg \cite{Stohr1972,Geng2014,Geng2015,Wu2015,Wu2018,Zhang2019} 
or Ti \cite{Minonishi1982,Numakura1986}, regardless of the pyramidal glide plane.

The critical resolved shear stress (CRSS) necessary to activate glide of \hkl<c+a> dislocations strongly depends on the temperature 
but is always much higher than the CRSS of \hkl<a> dislocation glide \cite{Akhtar1973,Gong2015}, 
thus explaining the strong plastic anisotropy of zirconium single crystals.
An increased activity of \hkl<c+a> slip is observed in irradiated zirconium alloys, 
as the hardening induced by irradiation defects appears stronger on \hkl<a> slip systems than on \hkl<c+a> pyramidal slip \cite{Long2015b,Long2016},
thus partly compensating for the friction difference between the two types of dislocations.

Although the mobility of \hkl<c+a> dislocations is a key ingredient of the plastic anisotropy of zirconium, not much is known on this mobility 
besides the slip plane, the dislocation ability to cross-slip and their tendency to align in a specific direction as mentioned above.
\Insitu{} TEM straining experiments are a valuable tool to study dislocation mobility. 
Such experiments in zirconium have already revealed some key features of \hkl<a> dislocations
\cite{Clouet2015,Long2015b,Caillard2015,Drouet2016,Caillard2018,Gaume2018}. 
But no such \insitu{} observations have been performed for the glide of \hkl<c+a> dislocations in zirconium. 
The difficulty arises from the necessity to have grains well oriented to activate \hkl<c+a> slip. 
The aim of the present article is to present such experiments.
\Insitu{} TEM straining experiments have been performed in two different zirconium alloys, Zircaloy-4 
and pure zirconium, as previous experimental studies on polycrystals \cite{Jensen1972} have shown 
a difference in the strain accommodation along the \hkl<c> axis between Zircaloy-4 and pure zirconium.
The results of these \insitu{} experiments are presented below.

\section{Experimental procedure}

\subsection{Materials}

Two different zirconium materials have been chosen for this study: zirconium sponge, which is essentially pure zirconium,  
and Zircaloy-4 in the recrystallized metallurgical state, which is often considered as a model material for zirconium alloys.
Their chemical compositions are given in Table \ref{tab:Zr_composition}.

\begin{table}[!tbh]
	\caption{Chemical composition (wt.\,\%) of zirconium materials used in this study.}
	\label{tab:Zr_composition}
	\centering
	\begin{tabular}{cccccc}
		\hline
			& Sn & Fe & Cr & O & Zr \\
		\hline
		Pure Zr & - & 0.019 & - & 0.024 & bal. \\
  		Zircaloy-4 & 1.32 & 0.215 & 0.108 & 0.125 & bal. \\
		\hline
	\end{tabular}
\end{table} 

\subsubsection{Pure zirconium}

Pure zirconium results from the Kroll process in the form of a 3\,mm thick plate. 
The raw zirconium sponge has first been arc-melted and cold-rolled with intermediate recrystallization treatments during 3 hours at 580\,$^{\circ}$C. After the last cold-rolling step, the same recrystallization treatment has been applied followed by a tensile test up to a strain of 1.5$\%$. The purpose of this straining was to reach critical strain annealing phenomenon in order to obtain large grains \cite{Chaubet2001,Zhu2005}. This straining was followed by an annealing at 840\,$^{\circ}$C during 24 hours. After this treatment, the mean grain size was 70\,$\mu$m with a large dispersion around this value. The material texture exhibited \hkl<c> axis mainly along the normal direction (ND) of the plate. Small samples, 3\,mm long and 1\,mm wide, were taken out of the plate with the long direction along ND, to activate \hkl<c+a> slip. 

\subsubsection{Zircaloy-4}

Recrystallized Zircaloy-4 is used for light water reactor applications as guide tubes. 
These tubes have a thickness smaller than 1\,mm and exhibit a strong texture with the \hkl<c> axis of the grains close to the radial direction (RD) of the tube, in the AD-RD (AD: axial direction) plane. 
Such products are not suitable to study easily the activation of pyramidal \hkl<c+a> slip, as one has to pull the grains along directions close to the \hkl<c> axis which are mainly along the radial direction of the thin tube. Because of the small thickness of the tube along the radial direction, it is not possible to machine small tensile test samples in this direction. This is the reason why an intermediate product of the fabrication process, called TREX for Tube Reduced Extrusion, has been used. This is a thick tube, with diameter of the order of 55\,mm and thickness of the order of 10\,mm. This material exhibits equi-axed grains with mean size of 6\,$\mu$m and a low initial dislocation density, only of the order of $10^{11}$\,m$^{-2}$.
The \hkl<c> axis of the grains are mainly oriented in the RD-TD (TD: transverse direction) plane with the highest pole density along TD \cite{Tournadre2012,Gharbi2015}. Small dog-bone specimens, 11.5\,mm long and 2.3\,mm wide, were machined by electrical discharge, with the tensile direction along TD. Therefore, thanks to the texture, a large number of grains are expected to have their \hkl<c> axis along the tensile direction.

\subsection[In situ TEM straining experiments]{\Insitu{} TEM straining experiments}

Pure zirconium and Zircaloy-4 samples have been mechanically polished down to 0.1\,mm thick and electropolished to make a small hole, using a mixture of 90$\%$ of ethanol and 10$\%$ of perchlorate acid at $-10$\,\degree{}C. The thin area around the hole is only few hundreds nanometer thick and is therefore suitable for TEM observations. 

\Insitu{} straining experiments were performed at room temperature on both materials using a GATAN  cooling and straining TEM sample holder on a JEOL 2010 operating at 200\,kV.
The thin foils were pasted on two ceramic rings or on copper grids at each ends to be pulled by the TEM holder.
For Zircaloy-4, experiments have been also conducted at room temperature on a FEI Tecnai operating at 200\,kV, using the commercial room temperature GATAN straining specimen-holder and the dog-bone TEM samples.
On these two specimen holders, one cross-head of the sample is fixed and the other cross-head is attached to a bar which displacement is controlled by a motor.
Typical displacements are 100\,$\mu$m. 

The experimental procedure is as follows. First, the orientation of various grains is analyzed until finding a grain with an orientation that satisfies two criteria: i) the \hkl<c> axis must be close enough to the tensile direction
(typically less than 10\,\degree{}) so that \hkl<c+a> slip will be activated, ii) the grain can be tilted, in the limit of the tilt range and with the restriction of single-tilt allowed by the TEM sample holder, so that a two beam diffraction condition with $\vec{g}=0002$ diffraction vector can be obtained. In this condition, the \hkl<a> dislocations disappear and only \hkl<c+a> dislocations are observed. The number of grains that meet these two conditions appears to be rather limited, despite the effort to select materials with adequate texture. 

Once the grains have been selected, the displacement of the cross-head is progressively increased until \hkl<c+a> dislocation motion is observed. The displacement is then maintained constant or slightly decreased and the dislocation motion is recorded using an ORIUS wide angle GATAN camera operating at ten images per second or a megaview III camera operating with 50 images per second. 
Supplementary video related to this article can be found at [INSERT DOI].

For some experiments, it has been possible to achieve extinction conditions for the \hkl<c+a> dislocations, by tilting the sample. This allows the determination of the exact Burgers vector of the dislocation. For most of experiments only the glide plane of the dislocation was deduced, without ambiguity, using two informations : i) the orientation of the slip traces and ii) the evolution of the distance between the two slip traces when tilting the thin foil. From this last measurement, the thickness of the thin foil can also be estimated. Furthermore, when cross-slip was observed, the Burgers vector of the dislocation has been deduced from the knowledge of the two gliding planes. 

\section{Results in recrystallized Zircaloy-4}

Besides the \insitu{} TEM straining experiments described below, 
some \postmortem{} observations on the same recrystallized Zircaloy-4 have been conducted 
after tensile tests to check that the behavior of \hkl<c+a> dislocations characterized \insitu{} 
in thin foils concurs with the dislocation microstructure obtained after straining macroscopic samples.
These \postmortem{} observations are described in \ref{sec:postmortem}.

During the \insitu{} experiments, the observation of \hkl<c+a> dislocation motion was always associated with the observation of twinning
which often occurred after \hkl<c+a> dislocation glide. 
The twinning system was \hkl[10-11]\hkl{10-1-2}
as expected in tension in zirconium alloys at room temperature \cite{Rapperport1960,Yoo1991,Viltange1985}. 
\ref{sec:twinning} describes an analysis of one twin event where it has been possible
to measure the twin propagation and thickening velocity.
In the following, we only focus on the description of \hkl<c+a> dislocation glide.

\subsection{Glide planes and dislocation orientation}

\begin{figure}[!bth]
	\centering
	\includegraphics[width=\linewidth]{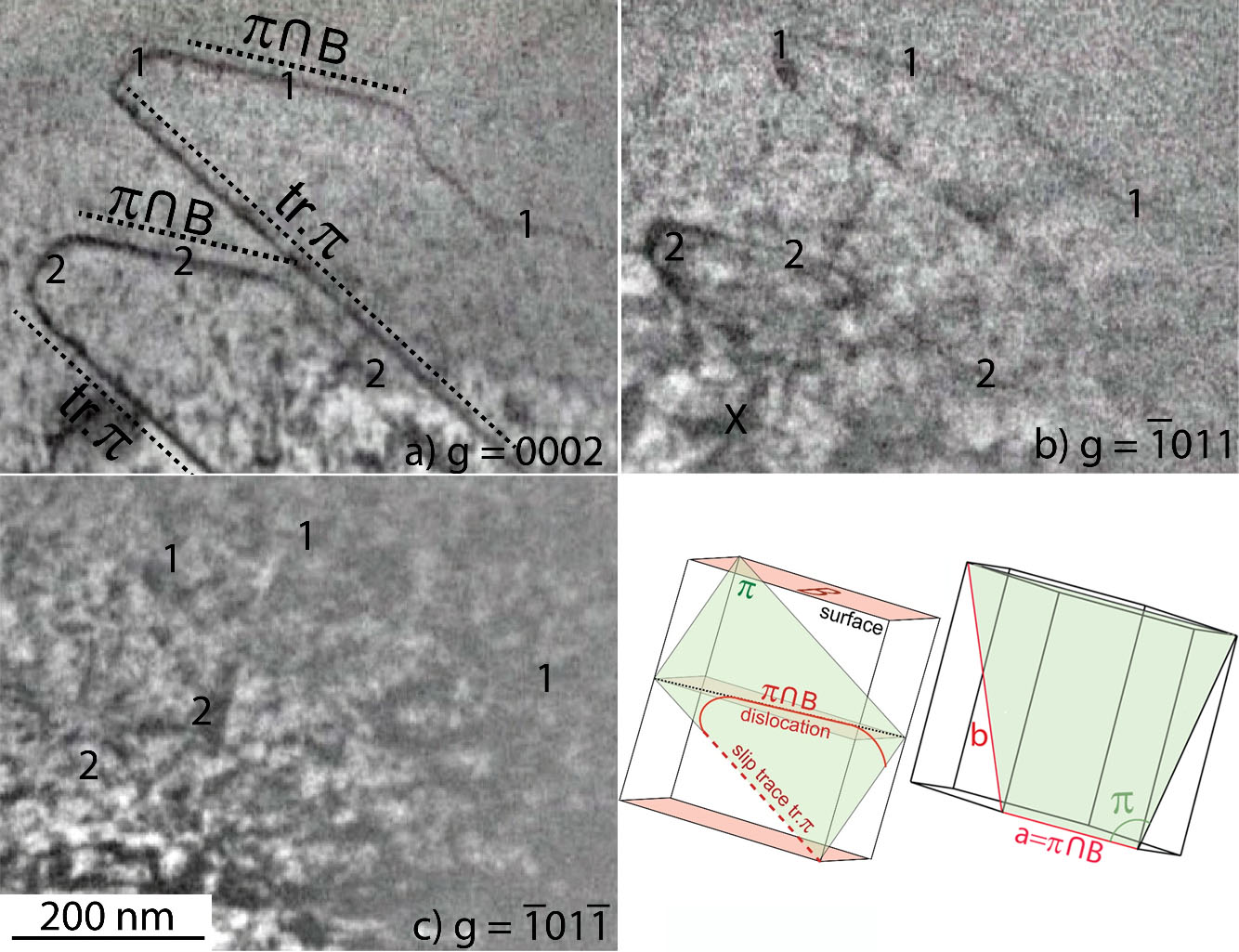}
	\caption{Zircaloy-4 strained in situ at room temperature. The dislocations are observed using a diffraction vector (a) $\vec{g}=0002$, (b) $\vec{g}=\bar{1}011$ and (c) $\vec{g}=\bar{1}01\bar{1}$. The tensile axis is vertical on the picture and close to the normal of the basal plane noted B. Two dislocations denoted as 1 and 2 glide in a \hkl(0-11-1) plane noted $\pi$. From the two possible Burgers vectors, the extinction shown in (c) gives $1/3\,\hkl[-1-123]$ as the Burgers vector. It is observed that the dislocations tend to be aligned along the \hkl[-1-120] direction noted $\pi \cap \textrm{B}$, which is the intersection of the first-order pyramidal glide plane and the basal plane.	
	}
	\label{fig:extinction}
\end{figure}

\begin{figure}[!bth]
	\centering
	%\includegraphics[width=0.65\linewidth]{figs/traces_py1_v3_1.png}
	%\hfill
	%\includegraphics[width=0.30\linewidth]{figs/traces_py1_v4_2.png}
	\includegraphics[width=\linewidth]{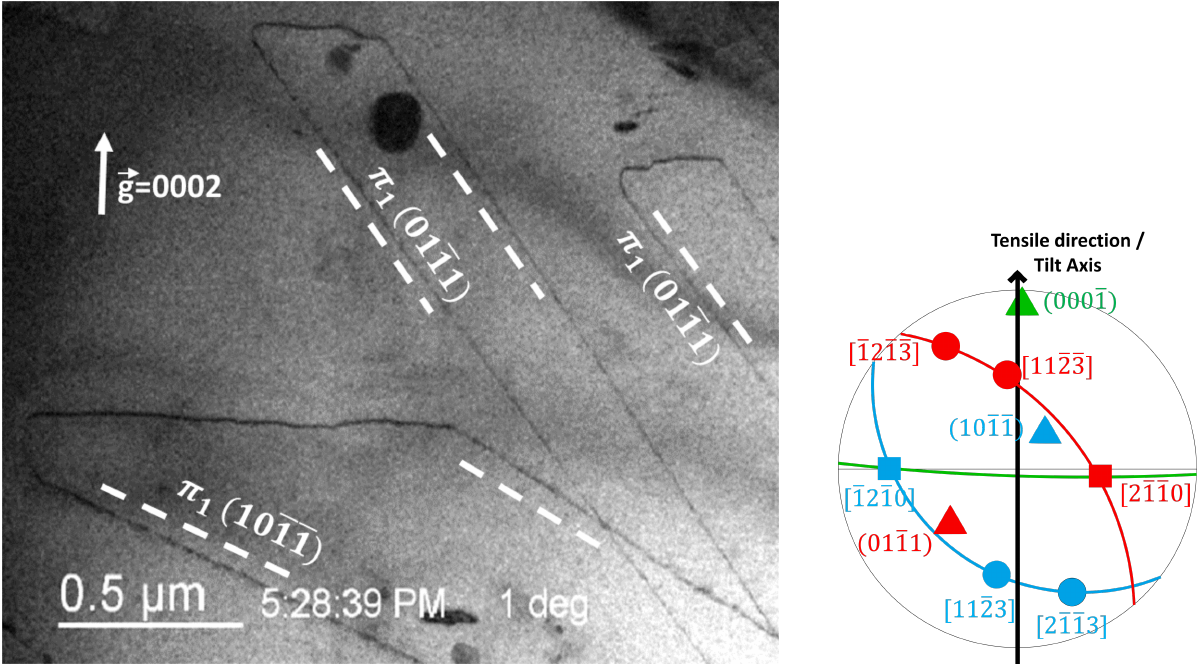}
	\caption{In situ TEM observations of \hkl<c+a> dislocations gliding in two different pyramidal planes.
	The corresponding slip traces are shown with dashed white lines. 
	For both slip systems, the dislocations are aligned along a \hkl<-2110> direction: 
	the \hkl[2-1-10] direction for the two short horizontal dislocations gliding in \hkl(01-11) 
	(slip plane in red in the stereographic projection) 
	and the \hkl[-12-10] direction for the long horizontal dislocation gliding in \hkl(10-1-1) (slip plane in blue).
	Possible $1/3\,\hkl<1-213>$ Burgers vectors are indicated by filled circles on the stereographic projection.}
	\label{fig:traces_py1}
\end{figure}

Gliding \hkl<c+a> dislocations in recrystallized Zircaloy-4 are shown in Figures \ref{fig:extinction} and \ref{fig:traces_py1}. 
In Figure \ref{fig:extinction}, it has been possible to determine exactly the Burgers vector $\vec{b}$ using different tilts: 
the dislocations are visible with diffraction vectors $\vec{g}=0002$ and $\vec{g}=\bar{1}011$, but invisible with $\vec{g}=\bar{1}01\bar{1}$,
thus showing unambiguously that $\vec{b}=1/3\,\hkl[-1-123]$. 
The dislocation glide plane could be determined from the direction of the slip traces on the surface and from the variations of the projected size of the curved dislocation and of its slip traces with the tilt angle: 
these $1/3\,\hkl[-1-123]$ dislocations 
glide in a \hkl(0-11-1) first-order pyramidal planes (Fig. \ref{fig:extinction}). 
All the slip traces analyzed during our \insitu{} tensile tests lead to a first-order pyramidal plane.
When several \hkl<c+a> slip systems were activated in the same grain (Fig. \ref{fig:traces_py1}),
they all correspond to the first-order pyramidal slip plane. 
In all grains where \hkl<c+a> slip was activated, the corresponding Schmid factors were higher than or equal to 0.36,
while the Schmid factors of \hkl<a> prismatic slip systems was smaller than 0.01.
As a consequence, no \hkl<a> dislocations was observed at this stage in grains where \hkl<c+a> slip is activated,
but \hkl<a> dislocations could be observed in the surrounding grains. 
As plastic strain increases, some \hkl<a> dislocations could sometimes also be observed in the same grains as \hkl<c+a> dislocations,
probably because of the complex mechanical load coming from the deforming surrounding grains, 
from the appearance of some cracks and from twinning, and also because of crystal rotation. 
But these \hkl<a> dislocations clearly appear latter than \hkl<c+a> dislocations. 
We could see some \hkl<c+a> dislocation nucleated on grain boundaries or on twins. 
But most of the time, \hkl<c+a> dislocations are coming from thick areas of the thin foil which could not be imaged.

In all these \insitu{} straining experiments (Figs. \ref{fig:extinction} and \ref{fig:traces_py1}), 
it can be noticed that the dislocation lines exhibit straight segments perpendicular to the $\vec{g}=0002$ diffraction vector. 
The \hkl<c+a> preferentially align in a direction defined by the intersection of their first-order pyramidal glide plane with the basal plane,
that is, along an \hkl<a> direction. 
This dislocation orientation is nearly edge with a character $\theta$ defined by
$\cos{(\theta)}=a/2\sqrt{a^2+c^2}$. 
For Zr ($c/a=1.593$), this leads to $\theta=75^{\circ}$.

\subsection{Dislocation mobility}

\begin{figure}[!b]
	\centering
	\includegraphics[width=\linewidth]{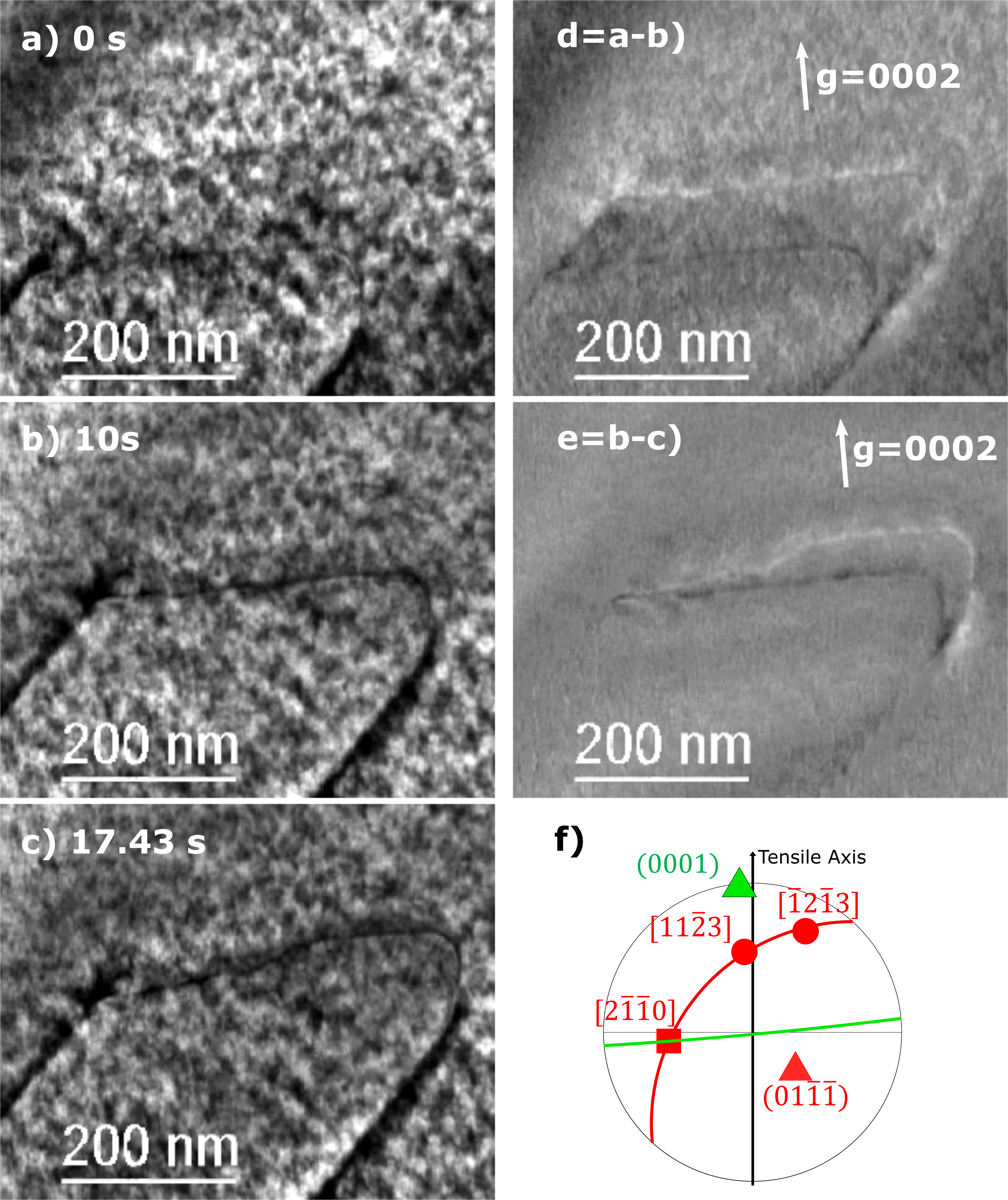}
	\caption{Glide of a \hkl<c+a> dislocation in the \hkl(01-1-1) pyramidal plane.  
	The same dislocation is observed at three different times: 
	the dislocation is almost straight in (a) and (b) with a line direction close to \hkl[2-1-10],
	whereas macro-kinks can be seen in (c).
	Figures (d) and (e) are obtained by subtraction respectively of images (a) and (b)
	and of images (b) and (c). 
	In these figures, the initial and final positions of the dislocation appear 
	with black and white contrast respectively.
	The dislocation Burgers vector has been determined to be $1/3\,\hkl[11-23]$
	among the two possible ones shown on the stereographic projection (f).
	See video as a supplementary material.
	}
	\label{fig:glissement_coin}
\end{figure}

\begin{figure}[!bth]
	\centering
	\includegraphics[width=\linewidth]{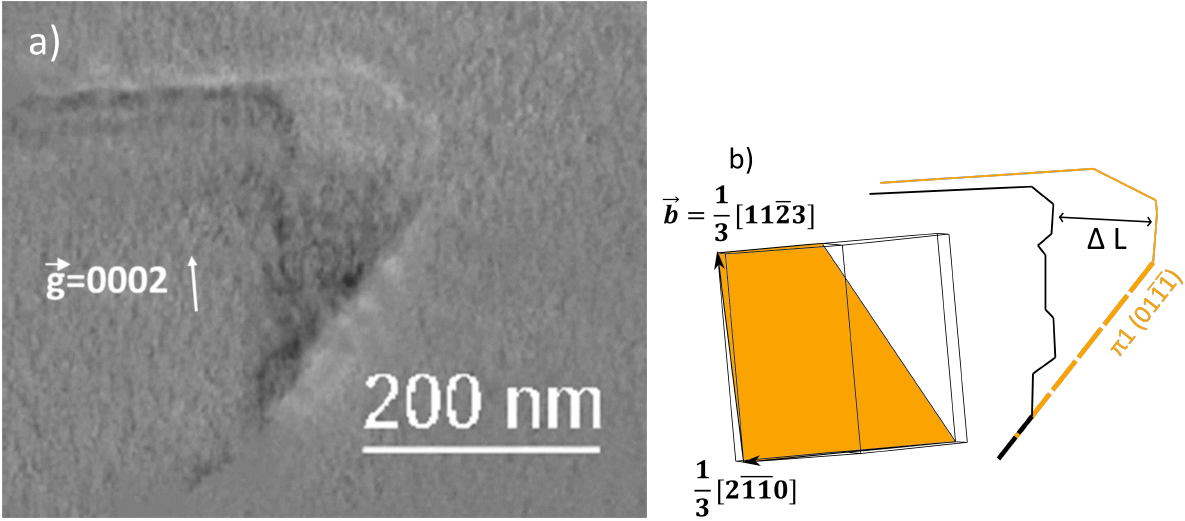}
	\caption{Jump of the screw part of the \hkl<c+a> dislocation shown in Fig. \ref{fig:glissement_coin}. 
	The image subtraction (a) corresponds to a time interval of 0.1\,s during which the screw part jumps a distance 
	$\Delta L\simeq 123$\,nm whereas the part aligned in the \hkl[2-1-10] direction moves slowly.
	The initial and final positions of the dislocation, as well as the associated slip traces,
	are sketched respectively in black and orange in (b).}
	\label{fig:glissement_vis}
\end{figure}

\begin{figure}[!bth]
	\centering
	\includegraphics[width=0.8\linewidth]{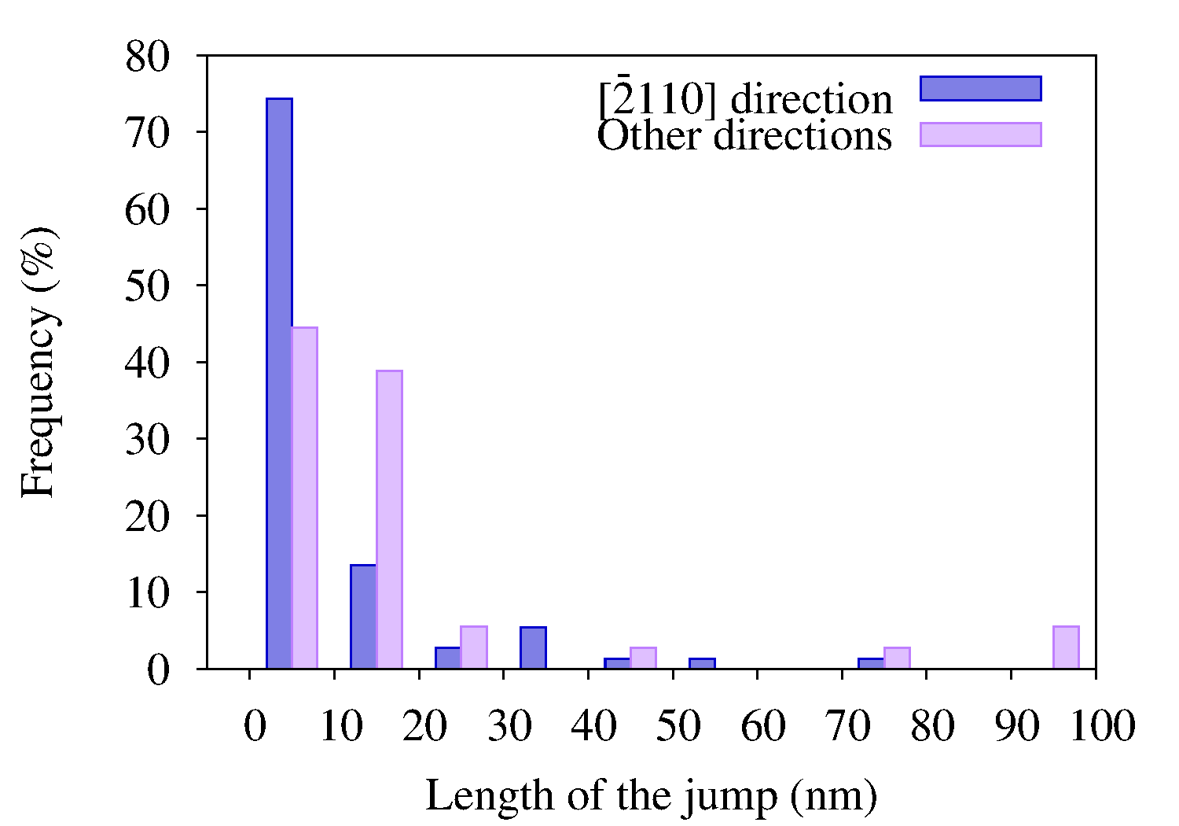}
	\caption{Jump length distribution for the \hkl<c+a> dislocation shown in Fig. \ref{fig:glissement_coin}. 
	The jumps have been measured on a time interval $\Delta t=0.1$\,s in a viscous motion regime and the whole analyzed sequence lasts 49.2 s.
	The blue histograms correspond to the jumps of dislocation segments along the \hkl[-2110] direction 
	and the purple ones to all other directions.}
	\label{fig:histogram}
\end{figure}

\begin{figure}[!bth]
	\begin{center}
		\includegraphics[width=\linewidth]{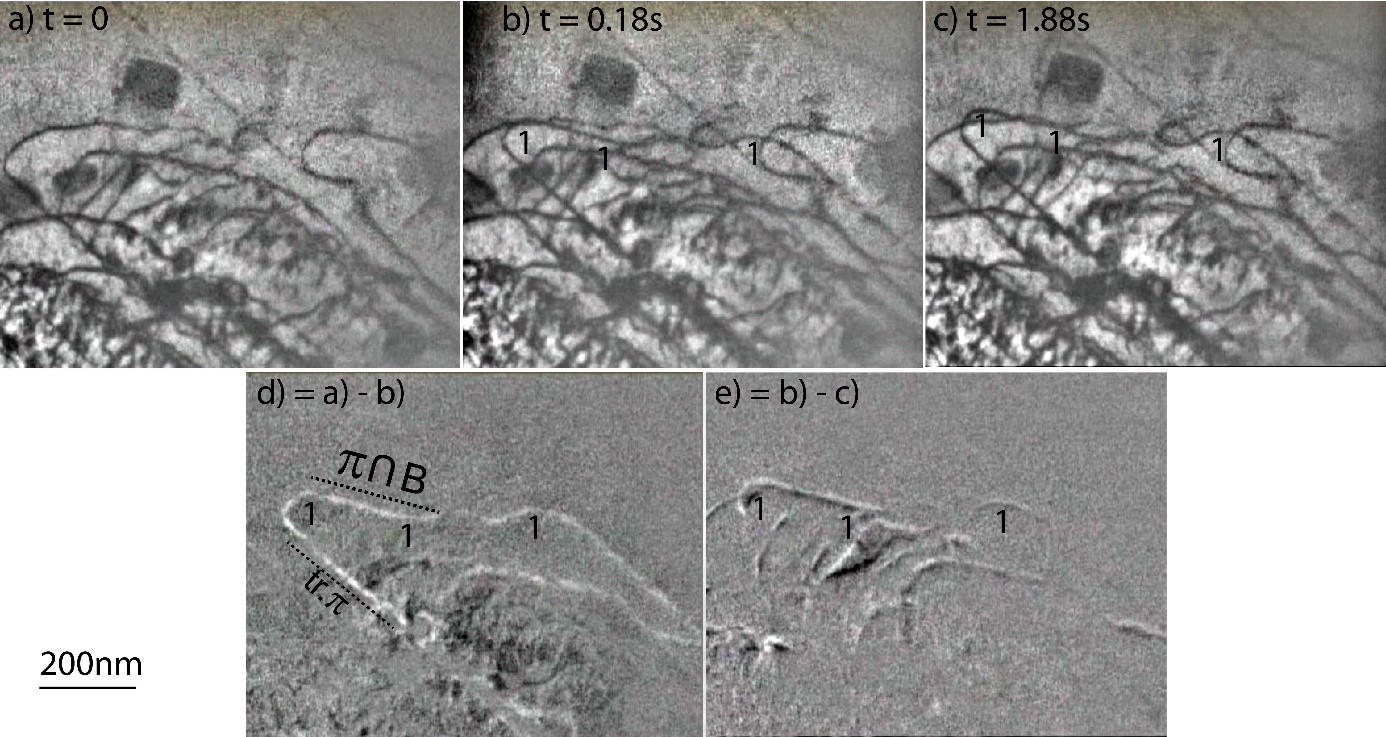}
	\end{center}
	\caption{Glide of \hkl<c+a> dislocations observed in the same zone as Fig. \ref{fig:extinction}.
	The dislocation labeled 1 arrived in the observation zone between images (a) and (b) 
	in a time interval $\Delta t=0.18$\,s. 
	Its straight portion aligned in the \hkl<a> direction noted $\pi \cap \textrm{B}$
	is then moving slowly between images (b) and (c) in $\Delta t=1.7$\,s.
	See video as a supplementary material.}
	\label{fig:glissement_rapide_coin}
\end{figure}

\Insitu{} observations of the dislocation motion in recrystallized Zircaloy-4 show that 
the \hkl<c+a> dislocations aligned along the \hkl<a> direction glide slowly in a rigid manner 
with occasionally the formation of macro-kinks.
This can be seen on Fig. \ref{fig:glissement_coin}, 
where the \hkl<c+a> dislocation aligned in the \hkl<a> direction is nearly horizontal 
and the slip traces are the diagonal lines.
Dislocation glide is better analysed on Figs. \ref{fig:glissement_coin}d and \ref{fig:glissement_coin}e which are obtained as subtraction 
of the same area observed at two different times so as to reveal the differences between the two images, 
and thus the dislocation motion in the time interval.
Between the initial time $t=0$ and $t=10$\,s, the nearly edge part of the \hkl<c+a> dislocation 
has kept its straight \hkl<a> orientation while gliding viscously a distance $\sim100$\,nm. 
At this time, this dislocation gets pinned in its middle with only one half keeping gliding 
in the following 7.4\,s.  This leads to the formation of a macro-kink. 
After the last frame shown in Fig. \ref{fig:glissement_coin}, the dislocation suddenly escapes from the observation zone, 
with a much faster motion.
Such a fast motion of the nearly edge part of the \hkl<c+a> dislocation has been captured 
in another \insitu{} straining experiment. 
In Fig. \ref{fig:glissement_rapide_coin}, one can see that the dislocation labeled 1 
has arrived suddenly in the observation zone in a time interval $\Delta t=0.18$\,s 
with a long part of the line aligned in the \hkl<a> direction 
noted $\pi \cap \textrm{B}$ on Fig. \ref{fig:glissement_rapide_coin}d. 
In the following 1.7\,s, the dislocation moves much more slowly (Fig. \ref{fig:glissement_rapide_coin}e).
Two mechanisms, leading either to a slow viscous glide or to a fast sudden motion, 
seem to control the mobility of this nearly edge part of the \hkl<c+a> dislocation.

Other orientations of the \hkl<c+a> dislocation, including the screw part, glide more rapidly with larger jumps than the viscous motion of the \hkl<a> orientation, 
as can be seen for instance on Fig. \ref{fig:glissement_vis}. 
The shape of these dislocations segments with other orientations is much more rounded and appears to be driven by minimization of the line tension.
Nevertheless, a lot of pinning points can be seen (Figs \ref{fig:extinction}, \ref{fig:glissement_vis} and \ref{fig:glissement_rapide_coin}),
showing that these segments are anchored on numerous localized point obstacles.

We have tried to evaluate the difference of mobility in the viscous glide regime between the part of the \hkl<c+a> dislocation
aligned along the \hkl<a> direction (nearly edge) and the other mixed-screw orientations for the sequence shown in Fig. \ref{fig:glissement_coin}. 
The glide distance $\Delta L$ between two frames ($\Delta t=0.1$\,s) is measured for each part of the dislocation (Fig. \ref{fig:glissement_vis}b).
If the dislocation does not move in this time interval, this is added to the waiting time of the dislocation.
The overall waiting time accounts for 70$\%$ of the sequence duration ($\Delta t=49.2$\,s). 
Only the time intervals corresponding to a dislocation motion are then analyzed, 
grouping measured glide distances by bins of width 10\,nm so as to build a histogram 
showing jump frequency as a function of glide distances 
both for the nearly edge and the other mixed orientations of the \hkl<c+a> dislocation (Fig. \ref{fig:histogram}).
It then clearly appears that there is a higher frequency of short jumps, smaller than 10\,nm, for dislocation lines along the \hkl<a> direction than for the other mixed-screw parts,
thus leading to a lower glide velocity of this near edge orientation. 
The large jump of 100\,nm long of the mixed-screw part shown in Fig. \ref{fig:glissement_vis} corresponds to the tail of the histogram.

\subsection{Cross-slip}

\begin{figure}[!tb]
	\centering
 	\includegraphics[width=\linewidth]{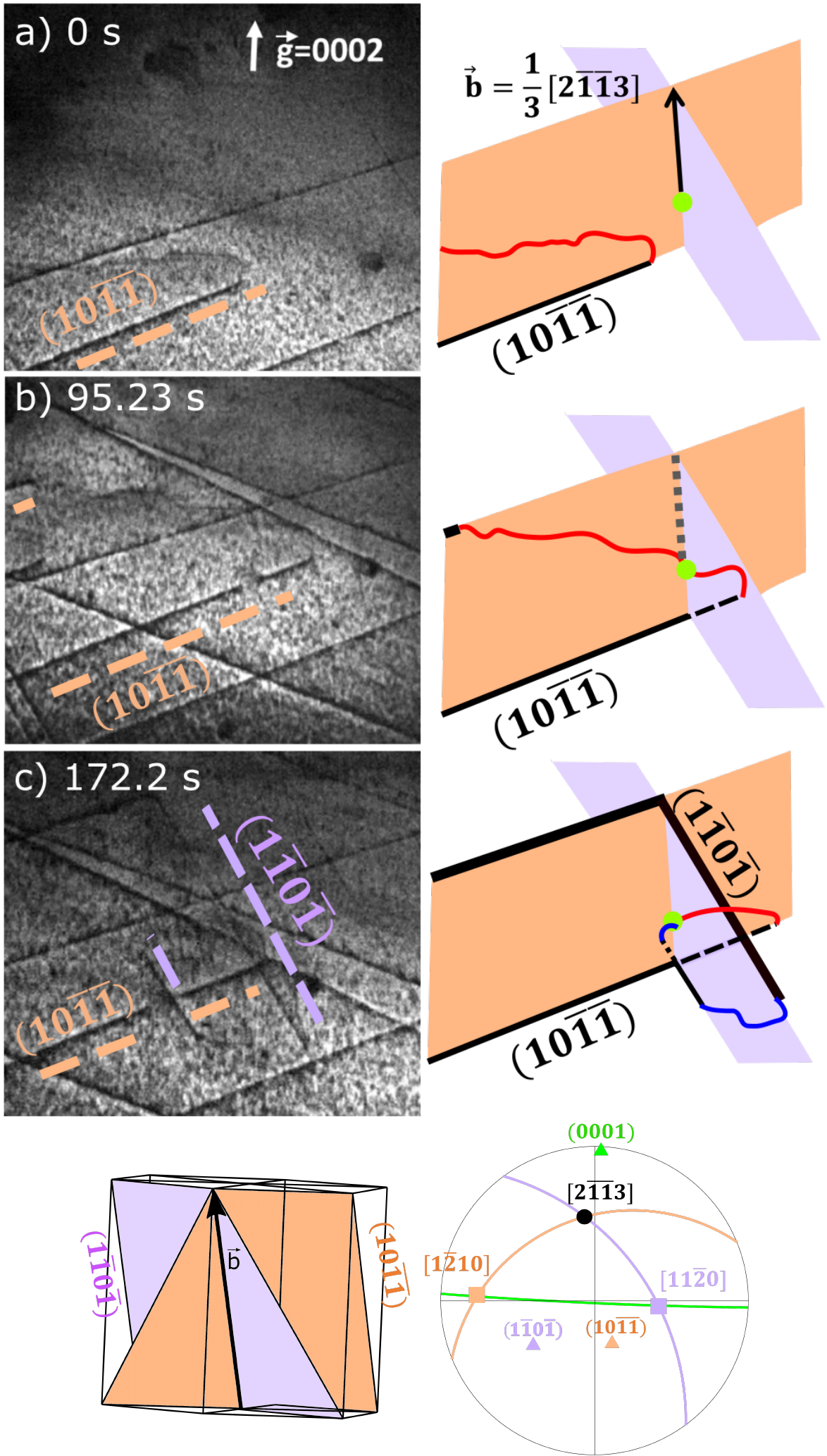}
	\caption{Dislocation cross-slip. 
	(a,b) The dislocation with a Burgers vector $1/3\,\hkl[2-1-13]$ is initially gliding in the \hkl(10-11) plane.
	Its interaction with an unknown defect makes it cross-slip. 
	(c) The dislocation pinned by this defect continues gliding in the initial \hkl(10-11) plane
	and in the cross-slipped \hkl(1-101) plane.
	On the right sketches, the dislocation is drawn with a red and blue line when gliding 
	respectively in its initial and crossed-slip plane. 
	Bold black lines are the slip traces on the thin foil surface.
	See video as a supplementary material.
	}
	\label{fig:cross_slip}
\end{figure}

During \insitu{} straining experiments, cross-slip of  \hkl<c+a> dislocations has been observed. One of these cross-slip events is illustrated on Fig. \ref{fig:cross_slip},
where the dislocation exhibits a light contrast and the slip traces a strong contrast. New slip traces appear with another orientation, coming from the initial slip traces. 
Both slip planes correspond to first-order pyramidal planes. When analyzing carefully the sequence to understand the geometry of the cross-slip event, it is found out that there must be an obstacle along the track of the dislocation. 
This obstacle could be a nano-hydride, a small precipitate or even a forest dislocation.
Because initially the dislocation is along the \hkl<a> direction, it is nearly edge and it cannot cross-slip. 
When the gliding dislocation meets the obstacle, which seems to be between the middle and the bottom of the thin foil, it is pinned. The part on the left continues to glide and bends towards the screw direction. When aligned along the screw direction, the dislocation cross-slips in the other first-order pyramidal plane. 

Four additional slip traces emerge between $t=0$ (Fig. \ref{fig:cross_slip}a) and $t=95$\,s (Fig. \ref{fig:cross_slip}b). 
These traces correspond to two \hkl<c+a> dislocations which have also slipped in two other first-order pyramidal planes, 
leading to two slip traces on the top surface of the thin foil and two on the bottom surface. 
The slip traces on the top and bottom surfaces are not parallel because of the wedge shape of the thin foil. 
We have checked that these additional \hkl<c+a> dislocations did not interact with the ones which have cross-slipped.

The process detailed here points out that \hkl<c+a> dislocations can cross-slip when they meet an obstacle 
and that the cross-slipped plane is also a first-order pyramidal plane.

\subsection{Relaxation}

\begin{figure}[!tb]
	\centering
	\includegraphics[width=0.7\linewidth]{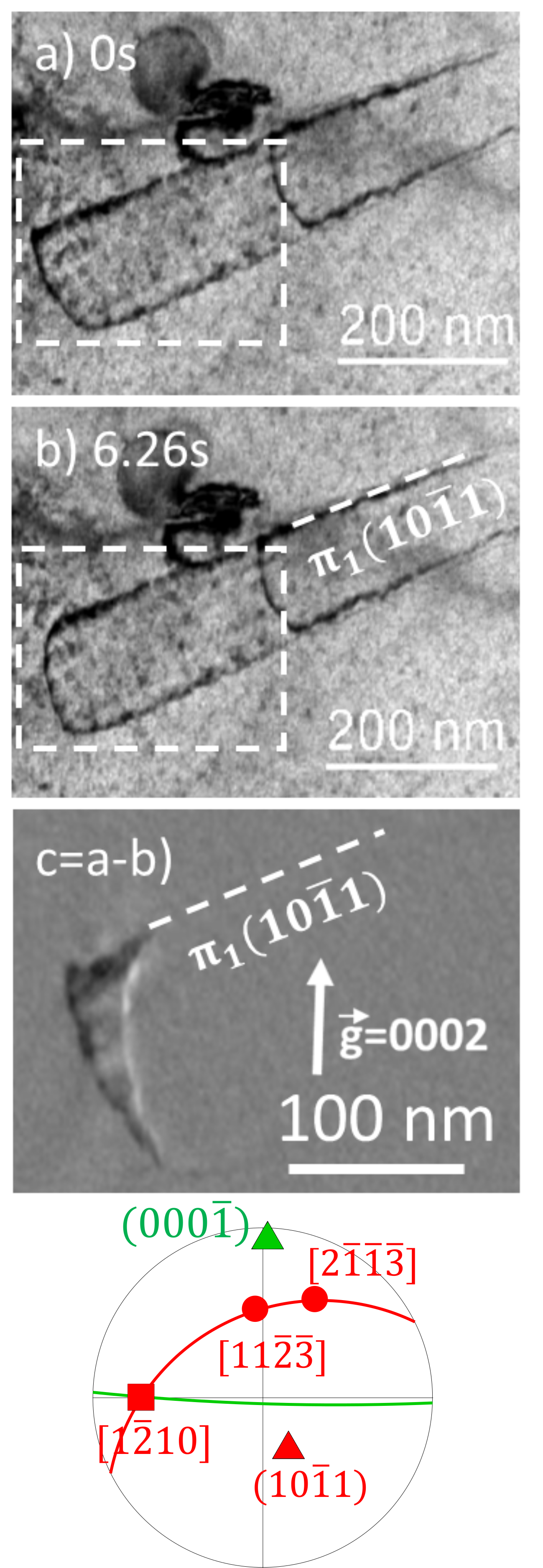}
	\caption{Relaxation of a \hkl<c+a> dislocation when the displacement 
	of the sample holder is slightly decreased.
	Between (a) the initial and (b) the final times, the dislocation on the left glides back 
	in its slip plane and partially removes the slip trace on one of the thin foil surface.
	This is better seen on the zoom-in differential image (c).
	The stereographic projection of the grain indicates that the gliding dislocation 
	is close to a screw orientation,
	its Burgers vector being either $1/3\,\hkl[11-2-3]$ or $1/3\,\hkl[2-1-1-3]$.
	See video as a supplementary material.
	}
	\label{fig:glide_back}
\end{figure}

When decreasing the stress applied on the specimen, by reducing the displacement of the cross-head from 120 to 100\,$\mu$m, 
a dislocation has been found to move back on its track (Fig. \ref{fig:glide_back}).
This dislocation relaxation has only been observed once in Zircaloy-4, but has also been observed in pure Zr. 
In the case of Zircaloy-4 shown on Fig. \ref{fig:glide_back}, 
it happens for a \hkl<c+a> dislocation which is far from the specific \hkl<a> orientation. 
When decreasing the applied stress, the dislocation moves back to align in a direction closer to the screw orientation, 
since screw dislocations have a lower line energy than dislocations with an edge character.
This relaxation of the \hkl<c+a> is therefore probably driven by the line tension, 
thus further showing that the lattice friction is not as strong for these orientations than for the \hkl<a> direction.
One also notices that the back motion of this \hkl<c+a> dislocation erases the slip traces on the thin foil,
proving that, at least in this particular case, dislocation glide is well confined in a single crystallographic plane.

\section{Results in pure Zr}

The same \insitu{} TEM straining experiments have been performed at room temperature on pure zirconium.  
Comparing these observations with the previous ones in Zircaloy-4 gives some insights on alloying effects 
on the mobility of \hkl<c+a> dislocations.

\subsection{Dislocation motion}

\begin{figure}[!b]
	\centering
	\includegraphics[width=\linewidth]{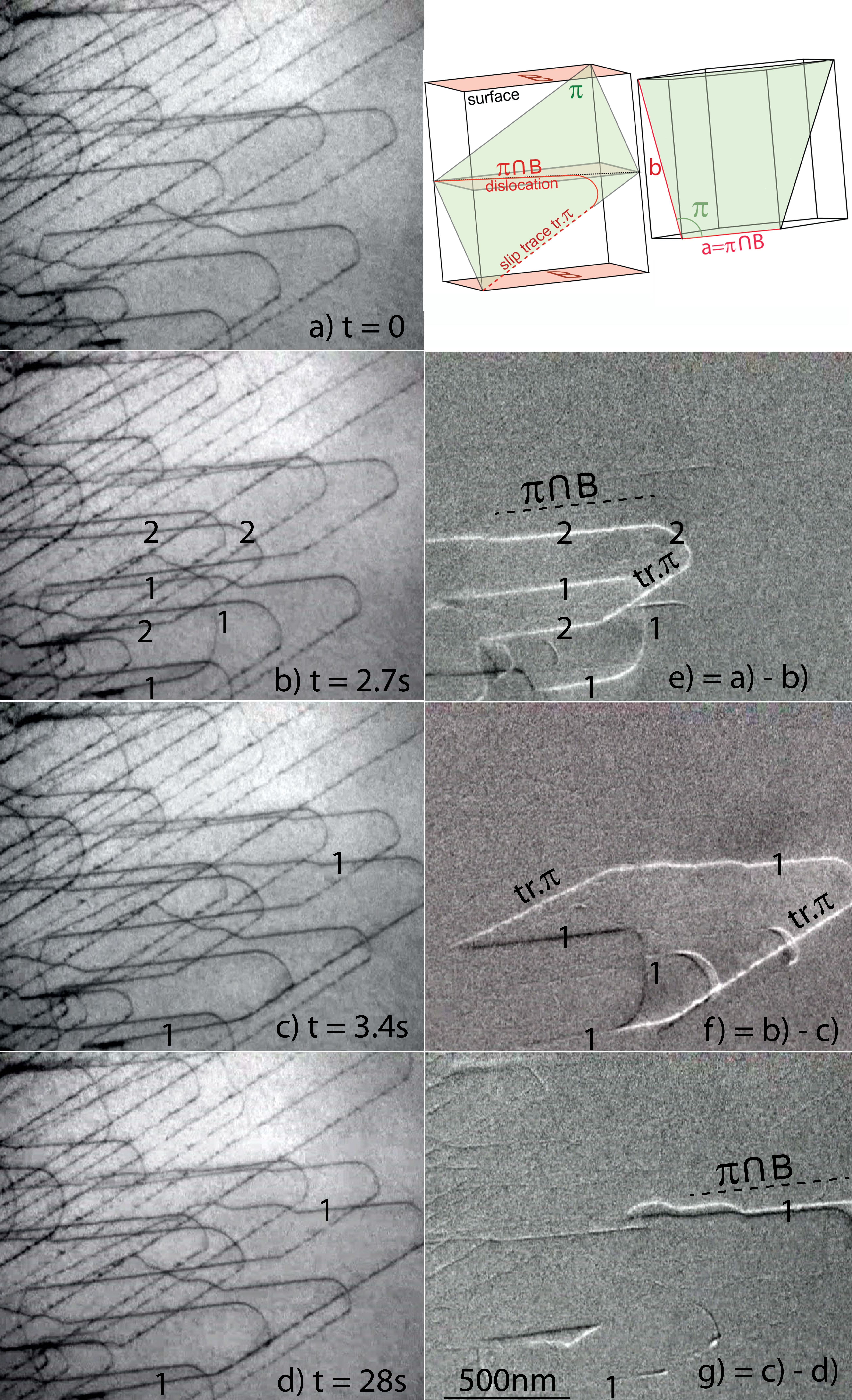}
	\caption{Pure Zr strained at room temperature. 
	The \hkl<c+a> dislocations are gliding in a pyramidal plane and are preferentially aligned in the \hkl<a> direction 
	defined by the intersection of the pyramidal glide plane with the basal plane.
	Two dislocations denoted as 1 and 2 appear between $t=0$\, and 2.7\,s: 
	the dislocation 1 pulls two straight \hkl<a> segments and ends with a curved line,
	while the dislocation 2 has slipped to the surface where it has left a slip trace  tr.~$\pi$.
	Between $t= 2.7$ and 3.4\,s, the upper and right parts of dislocation 1 unlock and jump to a new position
	while its lower straight section extends to the surface without moving.
	Between $t=3.4$ and 28\,s the upper straight part of the dislocation slightly glides
	but without any lateral motion of macro-kinks present on this dislocation.
	See video as a supplementary material.
	}
	\label{fig:zr_alignement}
\end{figure}

Like in Zircaloy-4, the slip planes of \hkl<c+a> dislocations have always been found 
in pure zirconium to correspond to the first-order pyramidal planes (Fig. \ref{fig:zr_alignement}).  
The dislocations also show some long rectilinear segments in the \hkl<a> direction defined by the intersection of the pyramidal glide plane with the basal plane.
Apart from this \hkl<a> orientation, the dislocations are curved, with a shape driven by the line tension.
One notices that in pure zirconium the straight portions are longer and better defined than in Zircaloy-4.  
This shape difference between both zirconium samples is a signature that the difference of lattice friction between this \hkl<a> direction 
and the other orientations of the \hkl<c+a> dislocations is stronger in pure zirconium than in Zircaloy-4. 
The presence of numerous solute elements in Zircaloy-4 slow-downs all \hkl<c+a> segments, 
thus reducing the mobility anisotropy of \hkl<c+a> dislocation. 
As pure zirconium contains very few impurities, this mobility anisotropy appears more important,
with an enhanced relative lattice friction acting against the glide of the straight \hkl<a> orientation.

\begin{figure}[!bt]
	\centering
	\includegraphics[width=\linewidth]{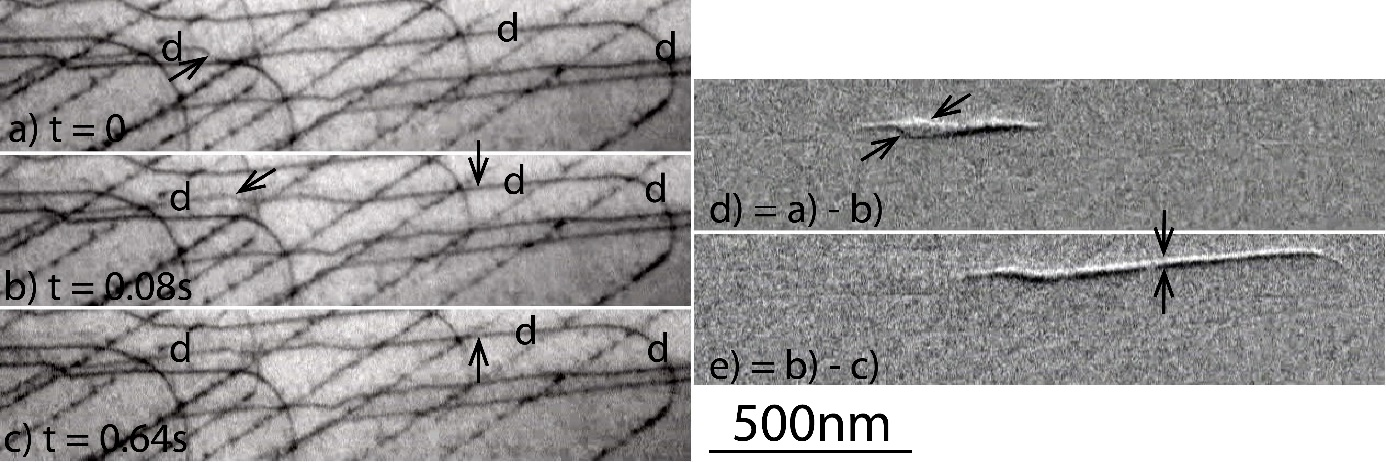}
	\caption{
	Detailed glide motion of the straight \hkl<a> part of the \hkl<c+a> dislocation in pure Zr (same zone as Fig. \ref{fig:zr_alignement}).
	The dislocation ``d'' is composed of segments lying along the intersection of the pyramidal glide plane and the basal plane.
	Between (a) and (b), a lateral motion of a macro-kink is observed (see inclined arrows)
	whereas the whole dislocation is gliding between (b) and (c)
	without any motion of the macro-kink (see vertical arrows in).
	See video as a supplementary material.
	}
	\label{fig:zr_mouvement}
\end{figure}

Like in Zircaloy-4, glide of the straight \hkl<a> orientation of \hkl<c+a> dislocations leads to the creation 
of macro-kinks (Fig. \ref{fig:zr_alignement}) in pure zirconium. 
These macro-kinks can either glide along the dislocation line (Fig. \ref{fig:zr_mouvement}d)
or remain fixed while the dislocation moves slowly (Fig. \ref{fig:zr_mouvement}e). 
Besides this viscous glide of the \hkl<a> orientation, one also observes some large jumps,
with therefore the same combination of slow and rapid motion as seen in Zircaloy-4.

A few cross-slip events have been also observed in pure zirconium.  
They appear, at first glance, less numerous than in Zircaloy-4, 
but a detailed statistical analysis would be needed to be able to really compare 
the occurrence of cross-slip in both materials.

\subsection{Friction stress}

Some relaxation observations have been also performed in pure zirconium. 
These experiments have been used to estimate the friction stress acting 
against the glide of the \hkl<a> orientation and of the other orientations, 
thus characterizing the mobility anisotropy. 
The friction stress originates from the dislocation core structure, 
like a possible non planar dissociation, and also from the dislocation interaction 
with the solute atoms present in the zirconium matrix.

\begin{figure}[!bt]
	\centering
	\includegraphics[width=\linewidth]{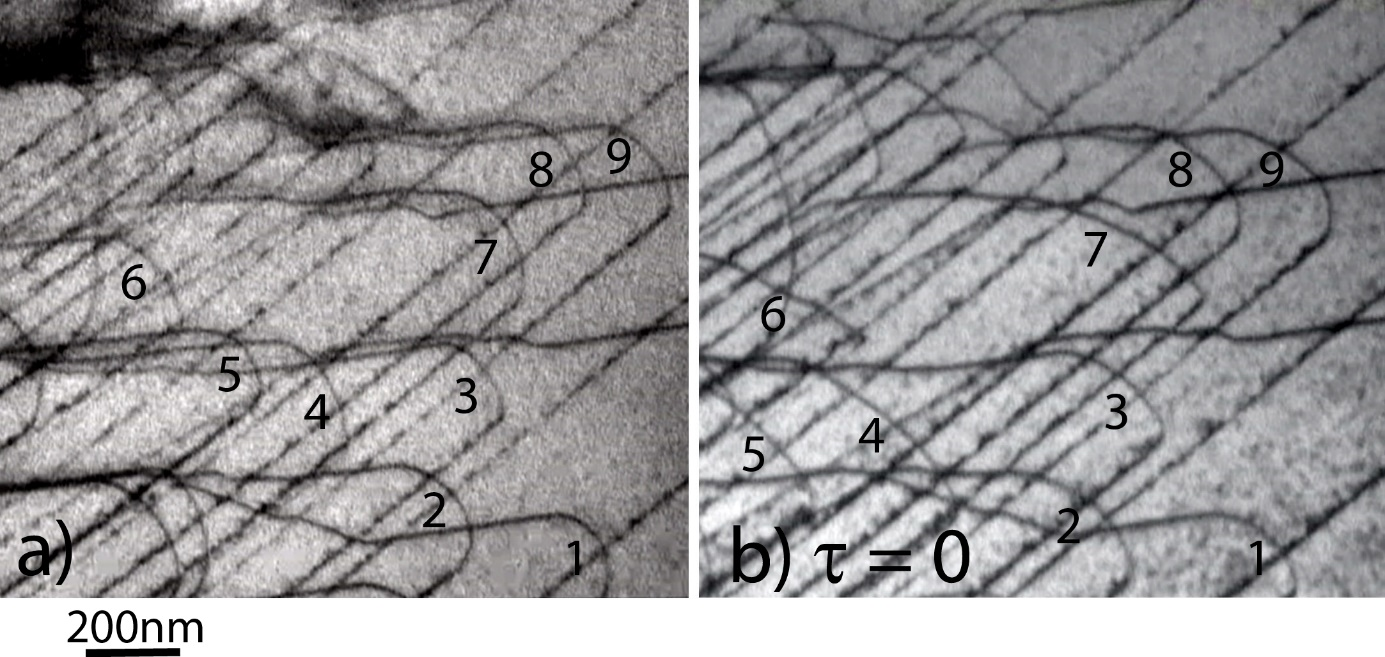}
	\caption{
	Relaxation of \hkl<c+a> dislocation in pure zirconium.
	The applied strain has been released between (a) and (b), leading to a back motion of the dislocations
	and an increase of their curvature radius.
	}
	\label{fig:zr_relaxation}
\end{figure}

When the applied stress $\tau_{\rm a}$ is removed, one observed that the curved parts of the dislocations 
are gliding back and that their curvature radius increases (Fig. \ref{fig:zr_relaxation}).
This variation of the curvature radius can be used to estimate the applied stress 
and the friction stress $\tau_{\rm f}$ acting against the glide motion of these curved segments. 
Using the program \textsc{Disdi} based on line tension calculations in anisotropic elasticity \cite{Douin1986}, 
one can deduce from the dislocation curvature the line tension stress $\tau_{\rm l}$. 
The shape of the loops 1 to 9 under an applied stress $\tau_{\rm a}$ shown on Fig. \ref{fig:zr_relaxation}a 
corresponds to a line tension stress $\tau_{\rm l}$ between 100 and 150\,MPa. 
When the stress is relieved ($\tau_{\rm a}=0$ on Fig. \ref{fig:zr_relaxation}b), 
we obtain a line tension stress $\tau_{\rm l}$ between 25 and 50\,MPa.
The applied stress is always in equilibrium with the line tension and the friction stresses, 
$\tau_{\rm a} = \tau_{\rm l} + \tau_{\rm f}$.
We therefore deduce a friction stress $\tau_{\rm f}=37\pm13$\,MPa acting again the glide of curved dislocation
and an applied stress $\tau_{\rm a}=160\pm40$\,MPa. 
This applied stress corresponds to the friction stress acting against the motion of the straight \hkl<a> parts of the dislocation, because the  line tension is zero for a straight dislocation,
thus confirming that this friction stress is much higher for this particular orientation.

\section{Discussion}

\subsection{Glide planes and cross-slip}

Only glide in first-order pyramidal planes has been observed in our experiments, 
both in pure zirconium and in Zircaloy-4.  
This agrees with previous TEM observations \cite{Akhtar1973,Woo1979,Merle1987,Numakura1991,Long2015a,Long2015b,Long2017,Long2018}
which invariably report the same glide plane for \hkl<c+a> dislocations. 
Only Long \etal{} \cite{Long2018} have shown that cross-slip in second order pyramidal plane is also possible, 
with nevertheless first-order pyramidal plane being the main glide plane. 
As the zirconium alloy studied by Long \etal{} contains 2.5\,wt.\% Nb, it is possible that 
niobium addition promotes the activation of this secondary slip system for \hkl<c+a> dislocations.

Cross-slip events between the two first-order pyramidal planes shared by a single \hkl<c+a> direction 
have been   observed in Zircaloy-4 and in pure zirconium in our experiments,
and also in Zr-2.5Nb alloy by Long \etal{} \cite{Long2015b,Long2017,Long2018}.
Cross-slip activation at room temperature may look surprising as the dislocation microstructure shows dislocations mostly aligned 
in a direction far from the screw orientation, which should prevent cross-slip. 
Nevertheless, cross-slip has been observed in our experiments only when the dislocation meets an obstacle, 
this obstacle forcing the dislocation to adopt locally a screw orientation
which can then cross-slip.
As \hkl<c+a> dislocations gliding in first-order pyramidal planes are thought to be dissociated in two partials separated by a stable stacking fault \cite{Rodney2017}, the obstacle should also help the dislocation constriction 
necessary for its propagation then in the cross-slipped plane, 
in the same way as the Friedel-Escaig mechanism operating in face-centered cubic metals \cite{Friedel1964,Escaig1968}.
As less obstacles are present in pure zirconium than in Zircaloy-4, one should expect cross-slip 
being less active in pure zirconium.  This looks in agreement with our observations, 
although, as mentioned earlier, a detailed statistical analysis would be needed to truly compare
cross-slip activity in the two materials.

\subsection[Straightening along <a> direction]{Straightening along \hkl<a> direction}

One key feature of the glide motion of \hkl<c+a> dislocations is the strong lattice friction
existing for the orientation corresponding to the intersection of the pyramidal glide plane 
with the basal plane, leading to long straight dislocations aligned along this \hkl<a> direction. 
This important friction exists both in pure zirconium and in Zircaloy-4, 
thus showing that it certainly derives from an intrinsic core property of this specific orientation.
As mentioned in the introduction, this behavior is not specific to zirconium, but can be found in other hcp metals.
In Mg, it has been observed that \hkl<c+a> dislocations gliding in second-order pyramidal planes 
may lock themselves by a non-conservative dissociation in the basal plane when their line direction belongs to this plane \cite{Stohr1972},
a mechanism which has been well described by elasticity \cite{Agnew2015} and atomistic simulations \cite{Wu2015}. 
For \hkl<c+a> dislocations gliding in first-order pyramidal plane, 
atomic simulations \cite{Numakura1990a,Numakura1990b,Wu2016a}
predict that the same locking mechanism operates, 
with a non conservative dissociation in the basal plane of the \hkl<c+a> in two partial dislocations separated by a basal $I_1$ stacking fault:
$1/3\,\hkl<1-213> \to 1/6\,\hkl<2-203> + 1/6\,\hkl<0-223>$.
Elastic calculations of the energy variation induced by this dissociation predict that 
it is also favorable in Zr \cite{Wu2016b}, thus potentially explaining the locking of the \hkl<c+a>
when they meet the basal plane.

However our \insitu{} experiments provide some clues that the lattice friction 
of this \hkl<a> orientation may have a different origin in zirconium.
Locking of edge \hkl<c+a> dislocations in Mg by a non conservative dissociation in the basal plane
has been correlated with the emission of basal prismatic loops with a $I_1$ stacking fault \cite{Stohr1972,Sandlobes2012}.
No such basal $I_1$ loops were detected in our \insitu{} TEM straining experiments, 
neither in pure zirconium nor in Zircaloy-4. 
As the emission of these loops is intimately related to the non conservative dissociation 
of \hkl<c+a> dislocations in the basal planes, this may be an indication that such a non conservative dissociation 
does not operate in zirconium at room temperature.  
Besides, contrary to magnesium \cite{Stohr1972},
\hkl<c+a> dislocations intersecting the basal plane are not completely sessile in zirconium 
but only suffer a reduced mobility compared to other orientations. 
Finally, if \hkl<c+a> dislocations lock themselves in the \hkl<a> direction 
by a non conservative dissociation, 
\ie{} by climb of partial dislocations in the basal plane, 
one would expect that the locking efficiency would depend on the arrest time. 
This would lead to dynamic strain ageing and to plastic instabilities 
which we did not observe in our \insitu{} experiments.
The large lattice friction of this \hkl<a> orientation of the \hkl<c+a> dislocation 
should therefore result from a different mechanism in zirconium.

This non conservative dissociation is not the solely possible mechanism leading to a lattice friction.
It has been also observed in Mg
that \hkl<c+a> dislocation may emit an \hkl<a> dislocation gliding in the basal plane
to leave behind a sessile \hkl<c> dislocation \cite{Stohr1972}. 
Nevertheless, this will lead once again to the permanent locking of the \hkl<c+a> dislocation
and not only to a reduced mobility. 
Besides, the extinction tests performed during our \insitu{} experiments (Fig. \ref{fig:extinction}) 
clearly indicate that the dislocations aligned along the \hkl<a> direction have a \hkl<c+a> and not a \hkl<c> Burgers vector. 
The possible dissociation $\hkl<c+a> \to \hkl<c> +\hkl<a>$ appears therefore not compatible with our observations,
unless the \hkl<c> and \hkl<a> are so tightly bound that they could not separate and should be considered as a single dislocation \cite{Wu2016a}.

Without going to the emission of a perfect basal \hkl<a> dislocation, 
Numakura \etal \cite{Numakura1990a,Numakura1990b} have shown that \hkl<c+a> dislocations
intersecting the basal plane can emit a single Shockley partial trailing a basal $I_2$ stacking fault. 
The corresponding dissociation, $1/3\,\hkl<1-213> \to 1/3\,\hkl<0-110> + 1/3\,\hkl<1-103>$,
does not require climb of the partial dislocations and 
leads to a non planar core with a stacking fault in the basal and in the pyramidal planes.
Atomistic simulations relying on a generic interatomic potential for hcp metals \cite{Numakura1990a,Numakura1990b}
show that this core is more stable than the planar core dissociated in the pyramidal plane, 
but that it recombines in this planar core under an applied stress
to easily glide then in the pyramidal plane. 
This non planar core could therefore explain the lattice friction of the \hkl<a> orientation.
Its ability to transform in a planar glissile core is also compatible with the jerky glide motion 
sometimes seen for this \hkl<a> orientation.

\subsection{Macro-kinks}

One characteristic of the gliding \hkl<c+a> dislocations when they are aligned in the \hkl<a> direction
is the presence of macro-kinks on these otherwise straight segments.
Recent \insitu{} TEM compression experiments performed in Mg \cite{Zhang2019} have evidenced the same macro-kinks 
on the straight portions of \hkl<c+a> dislocations, leading to ``stair-like'' shapes.
Although \hkl<c+a> dislocations are not gliding on the same pyramidal plane in Zr and Mg, 
their glide motion induces the formation of macro-kinks in both metals.
We can reasonably exclude that these macro-kinks are a by-product of a reaction with \hkl<a> dislocations
as extinction tests in TEM observations showed that they have the same \hkl<c+a> Burgers vector
as the whole dislocation lines. 
They could arise from a local change of configurations of the \hkl<c+a> dislocation, 
between an almost sessile core corresponding to the straight segments 
and a glissile core for the macro-kinks.
These macro-kinks are less numerous in pure zirconium than in Zircaloy-4,
leading to longer straight dislocations in the former case. 
Their interaction with the different solute atoms present in solid solution hinder their motion,
thus leading to a stronger pinning of these macro-kinks in Zircaloy-4 than in pure zirconium.

\subsection{Lattice friction}

Our line tension measurements indicate a friction stress of $160\pm40$\,MPa acting against glide of the straight part 
of \hkl<c+a> dislocations in pure zirconium.
This friction has both an intrinsic origin probably due to a complex core of the \hkl<c+a> dislocation
for the \hkl<a> orientation as discussed before and an extrinsic one arising from interaction with solute atoms.
This friction stress of the less mobile segments should correspond to the yield stress
of \hkl<c+a> pyramidal slip at room temperature.
Modeling with crystal plasticity their micro-cantilevers bending experiments, 
Gong \etal{} \cite{Gong2015} obtained for the same slip system a much higher value, $532\pm58$\,MPa,
in a commercial pure zirconium.
With a higher level of solute elements in commercial pure zirconium than in the pure zirconium 
used in our experiments, in particular oxygen concentration which is four times higher, 
this yield stress difference indicates a strong influence of alloying elements. 
This is further supported by the strongest pinning of gliding dislocations 
observed in Zircaloy-4 than in pure zirconium.

This reduced mobility of \hkl<c+a> dislocations in alloyed zirconium 
may appear at first glance in contradiction with the study of Jensen and Backofen \cite{Jensen1972} 
who observed that plastic strain along the $\hkl<c>$ axis was mostly accommodated by twinning below 300\,{\degree}C in pure zirconium
whereas only $\hkl<c+a>$ slip was activated in Zircaloy-4 with very few compressive twins.
Although solute elements further hinder glide of $\hkl<c+a>$ dislocations, this hardening contribution appears less important 
than the impact of the same solute elements on twinning with the suppression of twinning by impurities in zirconium \cite{Garde1973},
in particular oxygen which addition strongly decreases twin activity \cite{Viltange1985}.  

%\ECcom{Daniel, tu m'avais conseillé de lire les 2 Philos. Mag de Marc Legros sur Ti3Al. 
%Je l'ai fait (rapidement) et je ne vois pas trop quel lien je peux faire avec nos expériences.}

\section{Conclusions}

\Insitu{} TEM straining experiments performed both in pure zirconium and in Zircaloy-4
have shown that \hkl<c+a> dislocations glide exclusively in first-order pyramidal planes 
at room temperatures, with some cross-slip events between two pyramidal planes 
sharing the same \hkl<c+a> direction. 
A characteristic feature of the microstructure is that \hkl<c+a> dislocations align in the \hkl<a> direction 
defined by the intersection of their glide plane with the basal plane, 
leading to long straight dislocations.  
Dislocations segments aligned along this orientation experience a higher lattice friction than others, 
with a friction stress estimated to $160\pm40$\,MPa for the \hkl<a> orientation 
instead of $37\pm13$\,MPa for others in pure zirconium.
This \hkl<a> orientation exhibits two different glide motions, either viscous or jerky, 
and its frequent pinning in localized points leads to the creation of macro-kinks.
The same dynamic behavior of \hkl<c+a> dislocations has been observed 
in pure zirconium and in Zircaloy-4, except for a highest friction 
acting against the motion of all characters in Zircaloy-4 because 
of the numerous alloying elements in solid solution. 
The mechanisms controlling the mobility of \hkl<c+a> dislocations in zirconium 
show strong similarities with the ones operating in other hcp metals, 
in particular magnesium despite a different glide plane.

%\clearpage
\vspace{0.5cm}
\linespread{1}
\small

\textbf{Acknowledgments} -
The authors thank Framatome for providing the raw Zr sponge and the Zircaloy-4 TREX tube,
D. Nunes and S. Urvoy (SRMA/CEA) for the preparation of the pure Zr samples 
and B. Arnal (SRMA/CEA) for thin foils preparation.
This work is funded by the French Tripartite Institute (CEA-EDF-Framatome) through the GAINE project.

\appendix
\section{Post-mortem observations}
\label{sec:postmortem}

\begin{figure}[!bth]
	\centering
	\includegraphics[width=\linewidth]{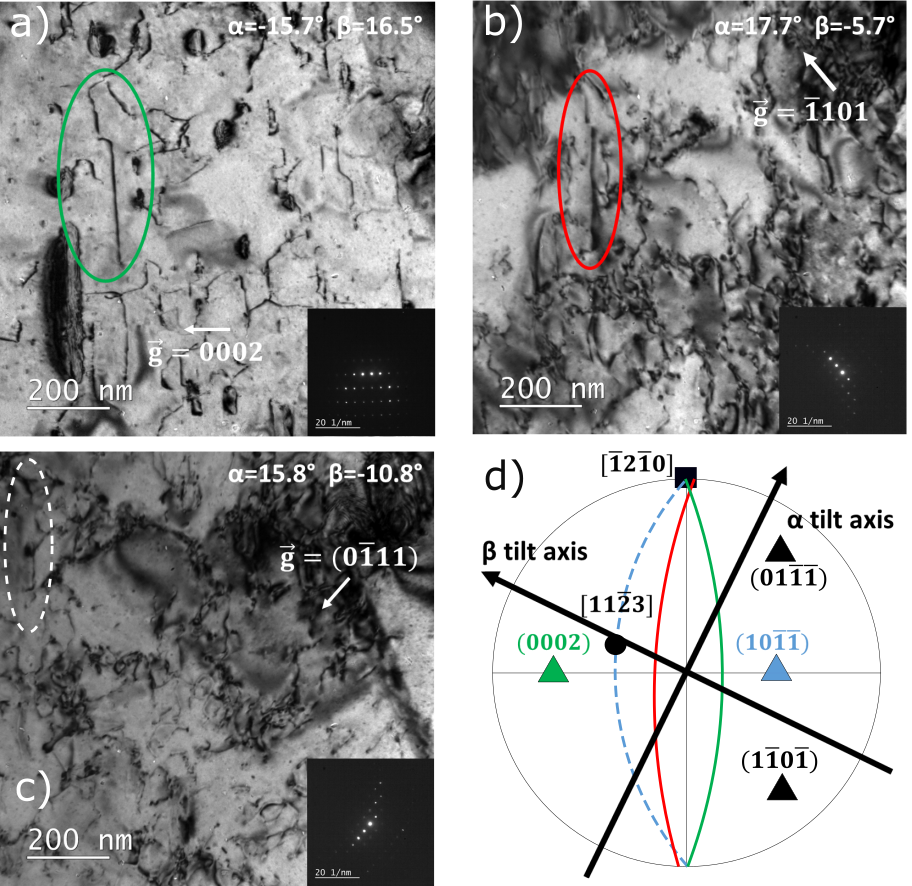}
	\caption{\Postmortem{} observation of \hkl<c+a> dislocations after conventional tensile test.  
		The region is observed with three different tilt angles $\alpha$ and $\beta$ 
		corresponding to the diffraction vectors (a) $\vec{g}=0002$, (b) $\vec{g}=\bar{1}101$ and (c) $\vec{g}=0\bar{1}11$.
		The color circles in (a) and (b) evidence the same \hkl<c+a> dislocation. 
		The white circle in (c) shows the position of the out of contrast dislocation
		since the diffraction vector extinguishes the dislocation. 
		The planes, observed edge-on in (a) and (b) for different tilts, containing the dislocation lines 
		are shown by the corresponding color lines
		in the stereographic projection (d) at zero tilt.
		Therefore, this dislocation with $1/3\,\hkl[11-23]$ Burgers vector 
		glides in \hkl(10-1-1) plane and
		is aligned along the \hkl[-12-10] direction.
	}
	\label{fig:post-mortem}
\end{figure}

Besides the \insitu{} TEM experiments described in the main text, 
some \postmortem{} observations have been carried out. 
The purpose was to check that the dislocation behavior observed \insitu{} in thin foils is typical of bulk plasticity.  
To this aim, conventional tensile tests on dog-bone Zircaloy-4 specimens taken out of the TREX tube with the tensile axis along the transverse direction have been performed at 350\,$^{\circ}$C up to a deformation of 4$\%$. 
Strained samples have been then observed with a JEOL 2100 TEM operating at 200\,kV. 

TEM observation with a $\vec{g}=0002$ diffraction vector allows the extinction of all \hkl<a> dislocations 
to image only dislocations with a \hkl<c> component.  
Such a \postmortem{} observation is shown in Fig. \ref{fig:post-mortem}.
Using other diffraction vectors, in particular $\vec{g}=0\bar{1}11$ for which the dislocation becomes invisible (Fig. \ref{fig:post-mortem}c),
we could evidence that this dislocation is not pure \hkl<c>, but is a \hkl<c+a> dislocation with $1/3\,\hkl[11-23]$ Burgers vector.
Stereoscopic analysis of this dislocation further showed that it is lying in the \hkl(10-1-1) first-order pyramidal plane.
This \hkl<c+a> is mainly aligned along an \hkl<a> direction corresponding to the intersection 
of its pyramidal glide plane with the basal plane.
All our \postmortem{} observations, like Fig. \ref{fig:post-mortem}, reveal the same properties for \hkl<c+a> dislocations,
in full agreement with the behavior observed during \insitu{} experiments.

\section{Twinning}
\label{sec:twinning}

\begin{figure}[!btp]
	\centering
	\includegraphics[width=0.8\linewidth]{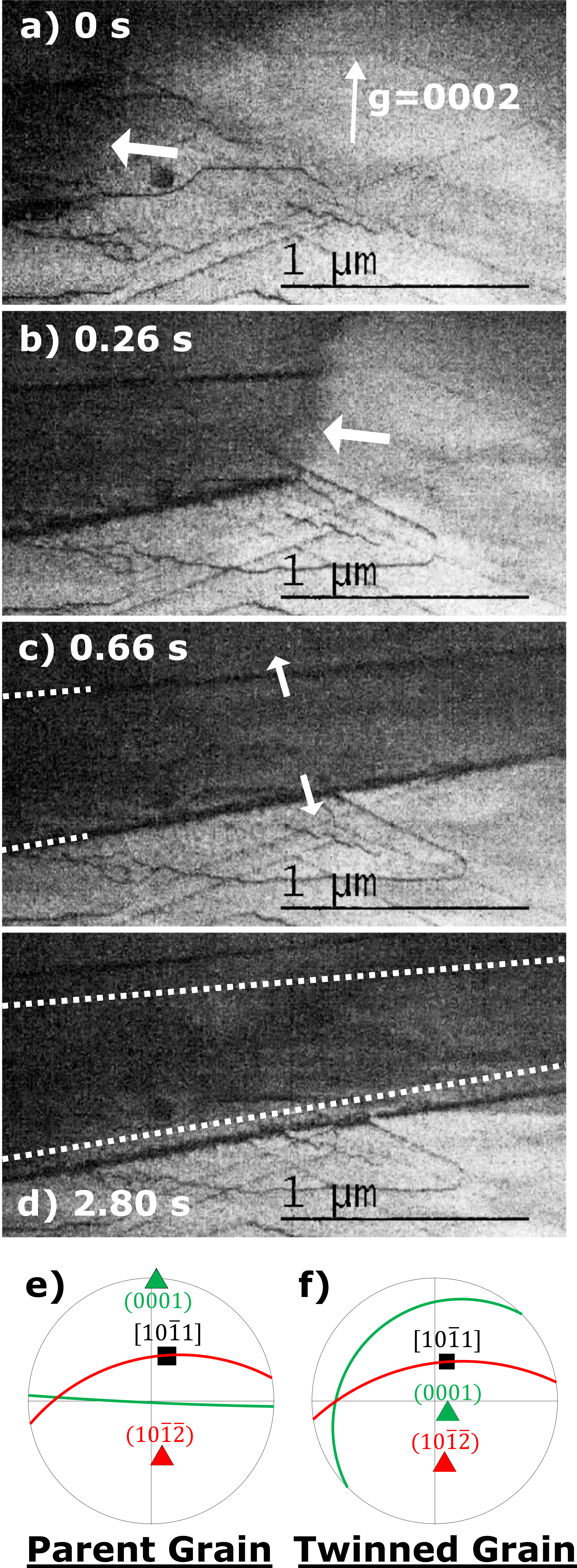}
	\caption{Nucleation and growth of a \hkl{10-1-2} twin. 
	The propagation of the twin tip can be seen between images (a) and (c), 
	while the thickening of the twin is observed, on a much longer timescale, 
	between images (c) and (d).
	The stereographic projection of (e) the parent and (f) the twined crystal 
	indicates that the twinning system is \hkl[10-11]\hkl(10-1-2).
	See video as a supplementary material.
	}
	\label{fig:twinning}
\end{figure}

Twining was observed \insitu{} in Zircaloy-4. In one case, the propagation of the twin and its thickening were slow enough to be analyzed (Fig. \ref{fig:twinning}).
The forefront of the twin appears to be very diffuse, without any distinct contrast, whereas the edge of the twin exhibits a strong contrast. The velocity of the propagation of the twin forefront was evaluated measuring the distance between each frame, taken every 0.1\,s. Because of the rapid velocity of the twin forefront, there are only 6 frames to measure the instantaneous velocity which were equal to 1.24, 1.60, 2.86, 4.27 and 6.03\,$\mu$m/s. 
Then, after the rapid propagation of the forefront of the twin, the twin is observed to thicken. 
The thickening velocity in the direction perpendicular to the twin interface is found to be equal to 33.3\,nm/s. 
The twin orientation with respect to the parent grain was analyzed:
this is a tension twin, corresponding to the twinning system \hkl[10-11]\hkl(10-1-2).

%\clearpage
\section*{References}
\bibliographystyle{elsarticle-num}
\biboptions{sort&compress}
\bibliography{soyez2020}

\end{document}